\documentclass[conference]{IEEEtran}
\IEEEoverridecommandlockouts
\usepackage[left=1.32cm,right=1.32cm,top=1.9cm,bottom=4.3cm]{geometry}
\usepackage{subcaption}
\usepackage{cite}
\usepackage{amsmath,amssymb,amsfonts}
\usepackage{graphicx}
\usepackage{textcomp}
\usepackage{tikz}
\usepackage{url}
\usepackage{tcolorbox}
\tcbuselibrary{breakable}
\usepackage{adjustbox}
\usepackage{multirow}
\usepackage{comment}
\usepackage{multicol}
\usepackage[affil-it]{authblk} 
\usepackage{amsmath}
\usepackage{algpseudocode}
\usepackage{amsmath,amssymb,amsfonts}
\usepackage{graphicx}
\usepackage{textcomp}
\usepackage{svg}
\usepackage{xcolor}
\usepackage{pgfplots}
\pgfplotsset{compat=newest,compat/show suggested version=false}
\usepackage{booktabs}
\usepackage{algpseudocode}
\usepackage[ruled, lined, longend, vlined,  linesnumbered]{algorithm2e}
\usepackage[utf8]{inputenc}
\usepackage{tikz}
 
\usepackage{amssymb}
\usepackage{pifont}

\def\BibTeX{{\rm B\kern-.05em{\sc i\kern-.025em b}\kern-.08em
    T\kern-.1667em\lower.7ex\hbox{E}\kern-.125emX}}
\begin{document}

\title{BotDetect: A Decentralized Federated Learning Framework for Detecting Financial Bots on the EVM Blockchains}

\author{Ahmed Mounsf Rafik Bendada}
\author{Abdelaziz Amara Korba}
\author{Mouhamed Amine Bouchiha}
\author{Yacine Ghamri-Doudane}

\affil[]{La Rochelle University, L3I, La Rochelle, France}

\maketitle
\begin{abstract}
The rapid growth of decentralized finance (DeFi) has led to the widespread use of automated agents, or bots, within blockchain ecosystems like Ethereum, Binance Smart Chain, and Solana. While these bots enhance market efficiency and liquidity, they also raise concerns due to exploitative behaviors that threaten network integrity and user trust. This paper presents a decentralized federated learning (DFL) approach for detecting financial bots within Ethereum Virtual Machine (EVM)-based blockchains. The proposed framework leverages federated learning, orchestrated through smart contracts, to detect malicious bot behavior while preserving data privacy and aligning with the decentralized nature of blockchain networks. Addressing the limitations of both centralized and rule-based approaches, our system enables each participating node to train local models on transaction history and smart contract interaction data, followed by on-chain aggregation of model updates through a permissioned consensus mechanism. This design allows the model to capture complex and evolving bot behaviors without requiring direct data sharing between nodes. Experimental results demonstrate that our DFL framework achieves high detection accuracy while maintaining scalability and robustness, providing an effective solution for bot detection across distributed blockchain networks.

\end{abstract}


\begin{tcolorbox}[breakable,boxrule=1pt,colframe=black,colback=white]
\scriptsize Paper accepted at the 2025 IEEE International Conference on Communications (ICC): Communication and Information System Security Symposium.
\end{tcolorbox}

\section{Introduction}

Decentralized Finance (DeFi) leverages smart contracts on blockchain platforms, such as Ethereum, Binance Smart Chain,  and Solana, to enable financial activities like lending, borrowing, and investing without intermediaries, relying on cryptoassets managed on distributed ledgers. The high level of automation in DeFi, combined with the significant volume of assets involved, attracts various types of automated agents, commonly known as bots \cite{b1}. Often deployed as externally owned accounts (EOAs), these bots operate autonomously to execute transactions without human intervention. By optimizing profits and facilitating essential functions such as arbitrage, liquidity management, and market making, they achieve a speed and precision beyond human capabilities. In the DeFi space, bots provide several benefits: they enable seamless and instantaneous trading, improve market efficiency by capturing arbitrage opportunities, and contribute to liquidity in decentralized exchanges and lending protocols. Through these roles, bots enhance the user experience and support the stability and growth of the ecosystem.

However, the widespread presence of bots also raises significant concerns for network security and stability, as many bots exploit the blockchain’s transparency to engage in manipulative behaviors. They can inflate transaction fees, manipulate markets, and capitalize on vulnerabilities within smart contracts ~\cite{b2}. These harmful activities have allowed bots to amass over a billion dollars in profits through practices such as front-running and sandwich attacks, impacting regular users and undermining the ecosystem's resilience. As DeFi continues to grow, the need for robust bot detection mechanisms becomes increasingly urgent to mitigate these risks and maintain trust within the network.

Currently, efforts to detect bots on the Ethereum blockchain remain limited. Transaction analysis tools such as MEV-inspect \footnote{\url{https://eigenphi.io/}} have been designed to analyze and identify profit-maximizing transactions, frequently performed by bots. However, their rule-based frameworks lack the adaptability required to capture the increasingly sophisticated and evolving behaviors of modern bots. As a result, these tools fall short in addressing the wide array of strategies deployed by bots in decentralized environments. 

Alternatively, data-driven approaches present a more flexible and responsive solution, better equipped to address bot behaviors' dynamic and adaptive nature. To the best of our knowledge, the only machine learning-based solution currently available in the literature is that proposed by Niedermayer et al. \cite{b1}. Their work introduces a pioneering machine learning framework for the detection of financial bots within the Ethereum platform. The authors begin by systematizing relevant scientific literature and gathering anecdotal evidence to construct a comprehensive taxonomy of financial bots, categorizing them into seven primary types and 24 subtypes. Subsequently, they develop a labeled dataset containing 133 human addresses and 137 bot addresses. With this dataset, they employ both unsupervised and supervised machine learning algorithms to facilitate bot detection within Ethereum. Nevertheless, a centralized machine learning approach necessitates data centralization for model training, a configuration that conflicts with the inherently decentralized architecture of blockchain and raises significant concerns regarding data privacy—an essential principle underpinning blockchain technology.

In response to these limitations, we propose an innovative, decentralized bot detection solution based on Federated Learning (FL) orchestrated through smart contracts on Ethereum. This approach respects Ethereum’s decentralized nature by allowing nodes to collaborate without directly sharing data, thereby preserving data privacy while enhancing blockchain security. Federated Learning’s decentralized structure also allows for greater adaptability and scalability across Ethereum’s distributed nodes, making it well-suited for the dynamic demands of bot detection. By leveraging federated learning, our solution advances bot detection within Ethereum while balancing efficiency, decentralization, and privacy.

The remainder of this paper is structured as follows: Section \ref{RL} provides an overview of related work. Section \ref{SOL} presents the methodology of the proposed framework. Section IV outlines the results of the performance evaluation, and finally, Section V concludes the paper.

\section{Background and Related Work} \label{RL}

Transaction analysis tools such as MEV-inspect\footnote{\url{https://github.com/flashbots/mev-inspect-py}} and Eigenphi\footnote{\url{https://eigenphi.io}} are capable of identifying MEV-associated transactions. Due to the time-sensitive nature of MEV activities, bots are commonly used to execute these transactions. These tools can thus be seen as bot detection systems, as they identify bots by tracing the sources of transactions related to MEV. Currently, these systems are limited to specific transaction patterns and follow a rule-based approach focused on a particular set of transactions coded within their system \cite{b1}.

Alongside these tools, various research initiatives have examined bot detection on Ethereum. Heuristic-based techniques have been employed to study bot activities, offering a cost-effective alternative to labor-intensive data annotation. For instance, graph-based heuristics have been used to explore different bot categories and to assess the frequency of private transaction usage in MEV activities~\cite{b3}. Time-based heuristics are valuable in distinguishing automated wallet actions by revealing distinct patterns in transaction timing~\cite{b4}. Additionally, metrics based on transactions and graph theory have been utilized to classify Ethereum wallets and uncover strategies such as sniping, where bots quickly acquire assets for profit~\cite{b5}. Furthermore, In \cite{b6}, the authors analyze the Ethereum and Binance Smart Chain (BNB) ecosystems of tokens. They introduce sniper bots, a new type of trader bot involved in on-chain activities. The study detects their presence and quantifies their activity. These studies establish a foundational approach to bot detection and suggest features that could be refined for an ML-based system with greater adaptability.

MEV bots, in particular, have garnered attention due to MEV’s implications for Ethereum's security and economy, as recent research has demonstrated~\cite{b2, b7}. Since timing is crucial in exploiting MEV opportunities, bots are typically programmed to identify these chances and execute transactions swiftly. Key studies have underscored the threats associated with MEV, focusing on real-time analyses of pending transaction data~\cite{b8}. Some work has also provided a systematized view of front-running on blockchains, drawing parallels with traditional finance and categorizing different forms of front-running based on the intent to exploit~\cite{b5}.

Federated Learning (FL) has gained significant traction in intrusion and attack detection, offering substantial benefits in terms of accuracy, privacy preservation, and scalability. By enabling distributed model training across multiple nodes, FL enhances detection accuracy by leveraging diverse data sources while keeping data localized on each device, thus preserving privacy. Studies such as \cite{b9,b10} have shown that FL can significantly improve intrusion detection capabilities without compromising sensitive information. Additionally, research like \cite{b11, b12} highlights FL’s capacity to distribute detection models efficiently, supporting real-time responses to emerging threats across interconnected environments. Blockchain technology has further advanced the potential of FL by providing a trustless, transparent, and secure framework for collaborative learning. Through smart contracts and consensus mechanisms, blockchain facilitates the decentralized training and updating of FL models, safeguarding model integrity and preventing tampering \cite{b10, b13}. Moreover, works such as \cite{b14, b15,  b16} underscore blockchain’s role in managing and verifying model updates across nodes, fostering an FL ecosystem that is both scalable and privacy-preserving.

Despite these advancements, FL’s potential remains largely untapped for security applications within blockchain, particularly for bot detection. Given FL’s benefits of decentralized training and privacy preservation, its application to bot detection holds considerable promise. This approach could harness local transaction data from multiple nodes to identify bot activities while safeguarding user privacy, ultimately strengthening the resilience and security of decentralized finance ecosystems. To our knowledge, none of the existing studies have explored using federated learning to identify bots in EVM blockchains.

\begin{figure*}
    \centering
    \includegraphics[scale=0.17]{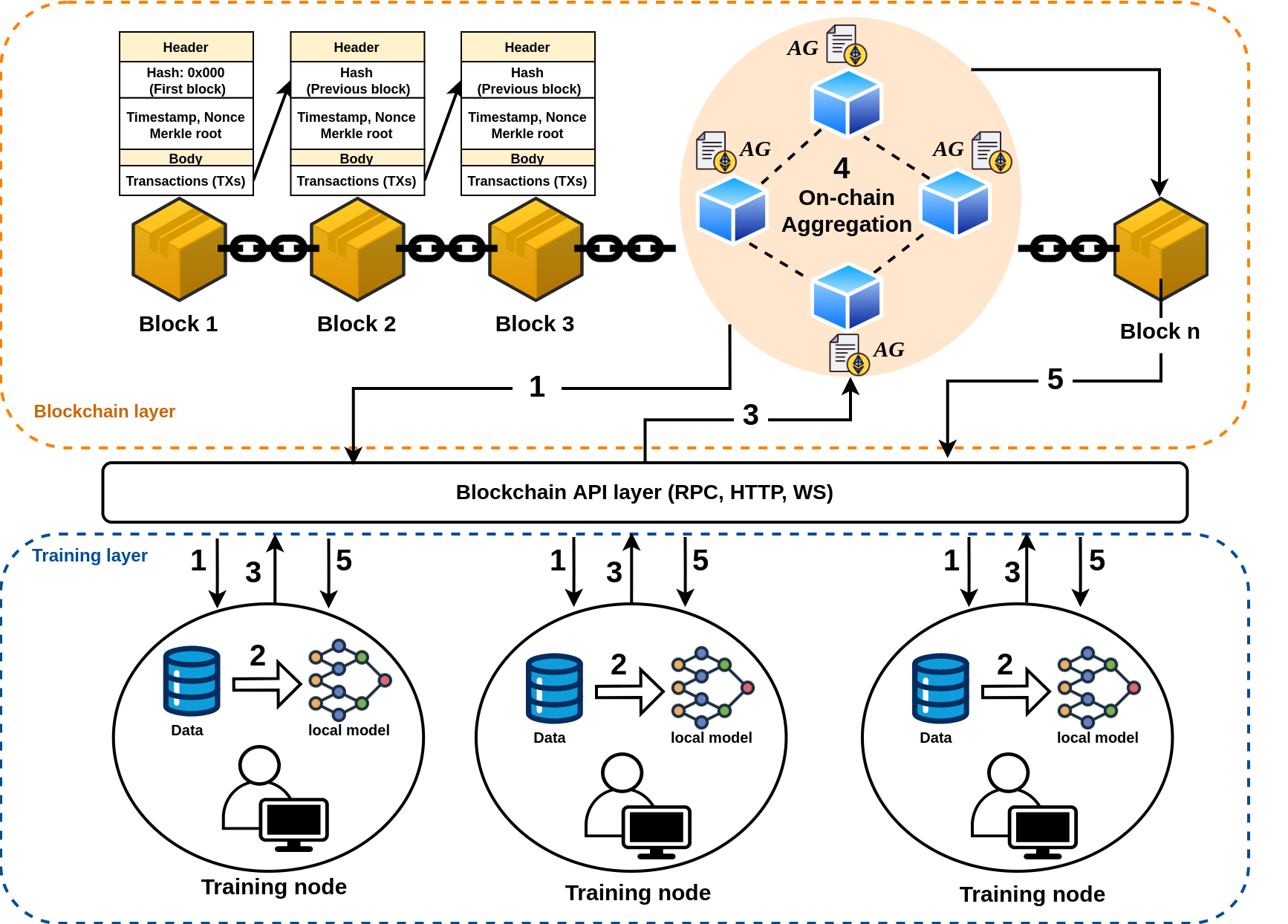}
    \caption{Workflow of the proposed framework: \textit{Step-1.} get initial global model; \textit{Step-2}. local training; \textit{Step-3}. submit update; \textit{Step-4.} aggregation; \textit{Step-5.} write global model and repeat.}
    \label{fig:archi3}
\end{figure*}

\section{Proposed Solution}\label{SOL}
We introduce an advanced bot detection framework employing deep learning (DL) models trained on transaction histories captured independently at each blockchain node. Leveraging federated learning, this approach ensures a substantial volume and diversity of data while preserving privacy, as data sharing between nodes is not required. We implement as decentralized orchestration of the federated learning process via smart contracts to blend with the inherently decentralized nature of blockchain. This design allows for distributed model sharing and aggregation, effectively bypassing the need for a centralized parameter server. We assume that participating nodes in the federated learning network are both reliable and cooperative. Notably, this federated learning occurs on a dedicated blockchain, isolated from the primary blockchain network to ensure streamlined operations. The subsequent sections elaborate on the system architecture and delineates the training process of the detection model.

\subsection{System Model} 
We define four key roles essential to the functionality and integrity of our decentralized federated learning (DFL) framework: Stakeholders, Validators, Training Clients, and Aggregators. Stakeholders ${S_i}$ include Traders, Liquidity Providers, Borrowers, Developers, and Protocol Founders. Validators ${V_i}$ are responsible for verifying and validating transactions through a Quorum-based Byzantine Fault Tolerance consensus protocol \cite{b17} in a permissioned setup. Stakeholders may serve dual roles, operating as blockchain nodes that contribute to network integrity through consensus and as training nodes that submit model parameters to the blockchain. This dual functionality, however, remains optional, allowing stakeholders to select roles based on their available computational and storage capacities.

Training Clients ${T_i}$ conduct local model training, while the Aggregator $AG$ smart contract merges weights from multiple local models to create a cohesive global model. Figure \ref{fig:archi3} provides an architectural overview of our DFL framework for bot detection.  Model training takes place off-chain at each node, with aggregation handled on-chain by the $AG$ smart contract to ensure decentralized coordination. This separation of model training from blockchain operations enhances modularity, supports a flexible federated setup, and preserves data privacy, as only local updates are shared, thus maintaining data diversity across nodes.

\subsection{Decentralized Federated Training}

To prepare the dataset effectively for training, it is essential to extract relevant features that enable the model to capture bot behavior. This extraction process is conducted off-chain, where data is gathered from the blockchain, processed, and converted into aggregate metrics. Off-chain processing enables complex calculations and aggregations that would be cost-prohibitive to perform on-chain, ensuring data richness and compatibility with the model. Our study utilizes the dataset produced in \cite{b1}, where transaction histories for each address are retrieved and analyzed, primarily focusing on outgoing transactions or token transfers, with some metrics derived from incoming transactions. The extracted features are categorized into four main types: address-based features, transaction-based features, function-call-based features, and event-based features. For further details on the features used, refer to \cite{b1}.

\color{blue}
\color{black}

The detection model is implemented as a fully connected Neural Network (FCNN), where each neuron performs a linear transformation on the input vector, followed by a non-linear activation function $f$. This process is represented as:

\begin{equation}
y = f(Wx + b)
\end{equation}

where $W$ is the weights matrix, $x$ is the input vector, $b$ is the bias term, and $f$ is the activation function. The number of nodes in the hidden layer in this work is set as follows:

\begin{equation}
\text{n} = \sqrt{m+k} + l
\end{equation}
where $n$ is the number of hidden nodes, $m$ is the number of input nodes, $k$ represents the number of output nodes and l is a constant from 1 to 10.

Each client node \( k \) independently and locally trains the detection model \( w_k \) based on its private dataset \( D_k \). The process begins with each client initializing its model using the current global model weights stored on the blockchain:

\begin{equation}
w_k^{(t,0)} = w^{(t)}
\end{equation}

The client then performs multiple epochs of Stochastic Gradient Descent (SGD) to minimize its local loss function, updating the weights iteratively as follows:

\begin{equation}
w_k^{(t,e+1)} = w_k^{(t,e)} - \eta \nabla F_k(w_k^{(t,e)})
\end{equation}

where \( \eta \) is the learning rate and \( \nabla F_k(w_k^{(t,e)}) \) represents the gradient of the local loss with respect to the model weights. After completing a predefined number of epochs, the client transmits the updated weights \( w_k^{(t)} \) to the blockchain, where they are securely recorded on-chain to ensure decentralized coordination.

Aggregation of these updates is managed by the smart contract \( AG \), which computes a average of the model weights submitted by each client to create the updated global model:

\begin{equation}
w^{(t+1)} = \frac{1}{K} \sum_{k=1}^K  w_k^{(t)}
\end{equation}

The aggregation can be weighted as follows:

\begin{equation}
w^{(t+1)} = \sum_{k=1}^K \frac{n_k}{N} w_k^{(t)}
\end{equation}

Here, \( n_k \) represents the size of each client’s local dataset, and \( N \) is the total data volume across all clients. This approach ensures proportional contribution from each client based on their dataset size, allowing for more fair aggregation. The aggregated global model is then made accessible to all clients and serves as the starting point for the next training round. This iterative process enables the DFL framework to securely and efficiently converge to a global model for bot detection, leveraging blockchain to maintain data privacy, integrity, and transparency. Algorithm \ref{algo:model_update_aggregation} details the on-chain aggregation process implemented by the $AG$ smart contract.

\begin{algorithm}[t]
\DontPrintSemicolon
\SetAlgoNlRelativeSize{-1}

\SetKwFunction{FSubmitModelUpdate}{submitUpdate}
\SetKwFunction{FAggregateModel}{aggregateModel}

\SetKwProg{Fn}{Function}{:}{}
\SetKwInOut{Input}{input}
\SetKwInOut{Output}{output}

\Input{Model update: \textit{update}, from: \textit{sender}}
\Output{Updated global model if aggregation threshold reached}

\Fn{\FSubmitModelUpdate{update, sender}}{
    \textbf{Require:} \textit{update}.length = \textit{aggregatedModel}.length\;
    modelUpdates[round][\textit{sender}] $\leftarrow$ \textit{update}\;
    trainers.push(\textit{sender})\;
    totalUpdates $\leftarrow$ totalUpdates + 1\;

    \If{totalUpdates $\geq$ $K$}{
        \FAggregateModel{}\;
    }
}

\Fn{\FAggregateModel{}}{
    newAggregatedModel $\leftarrow$ [0, 0, ..., 0]
    \Comment{length of aggregatedModel}\;
    \For{$i \in [0,1,...,\text{aggregatedModel.length}-1]$}{
        sum $\leftarrow$ 0 ; count $\leftarrow$ 0\;
        \For{$j \in [0,1,...,\text{totalUpdates}-1]$}{
            participant $\leftarrow$ trainers[j]\;
            sum += modelUpdates[round][participant][i]\;
            count += 1\;
        }
        newAggregatedModel[i] $\leftarrow$ sum / count\;
    }
   \textit{aggregatedModel} $\leftarrow$ newAggregatedModel\;
    round $\leftarrow$ round + 1\;
    totalUpdates $\leftarrow$ 0\;
    trainers.clear()\;
}

\caption{On-chain aggregation smart contract}
\label{algo:model_update_aggregation}
\end{algorithm}

\color{black}

\section{EXPERIMENTS AND RESULTS ANALYSIS}\label{SIM}
In this section, we first provide a brief description of the dataset used in this research. Next, we present and thoroughly discuss the detection and blockchain performance of the proposed system. Finally, we compare our approach with state-of-the-art (SOTA) methods.

\subsection{Data Preprocessing}
The dataset \cite{b1} used in this study comprises labeled data from the Ethereum blockchain, containing a total of 270 externally owned accounts (EOAs) labeled as either human or bot. Of these, 133 addresses are associated with human users, while 137 are identified as bots. The creation of this dataset involved expert manual annotation, with each address labeled according to observed activity patterns. This labeling distinguishes addresses controlled by humans from those operated by automated programs (bots). In addition to binary labels (human or bot), some bot addresses have specific labels that reflect their functional category, including common types of maximal extractable value (MEV) bots on Ethereum, such as arbitrage, sandwich, and liquidation bots. The characteristics of each address include transactional metrics based on transaction frequency, gas prices, and other key parameters, designed to capture behavioral patterns typical of bots.

Given the limited number of samples in the dataset and our need for additional examples to support federated training, we expand the dataset using the Synthetic Minority Over-sampling Technique (SMOTE) \cite{b18} with $\kappa=5$. We generated 5000 samples for each data class {Arbitrage, Liquidation, Sandwich, and Non-MEV} for the multiclassification datasets. For the binary classification dataset, we created 5000 examples for each category: bot and human. The dataset is then split into 70\% training and 30\% testing. The training data is further divided based on the number of federated clients, tested with 3, 4, and 5 clients by creating 3, 4, and 5 randomized subsets, respectively. Throughout this division, data balance was maintained for both bot/human detection and MEV attack classification scenarios.


\begin{table}[h]
\centering
\caption{Hyperparameter Settings of the Models}
\label{tab:params}
\begin{tabular}{@{}lcc@{}}
\toprule
\textbf{Hyperparameter}          & \textbf{Binary Classifier Values} & \textbf{Multiclass Values} \\ 
\midrule
Optimizer                        & Adam                              & Adam                       \\ 
Batch Size                       & 16                                & 16                         \\ 
Epochs                           & 50                                & 50                         \\ 
Learning Rate                    & 0.00001                          & 0.00001                    \\ 
Number of FL Iterations          & 5, 8, 10                         & 10, 12, 15                 \\ 
Number of Clients                & 3, 4, 5                          & 3, 4, 5                    \\ 
\bottomrule
\end{tabular}
\end{table}

\subsection{Detection Performance Evaluation}
The proposed system was trained and tested in the Google Colab cloud environment, using the Tensorflow Keras package for implementing both the local and federated learning models. Table \ref{tab:params} details the hyperparameter configurations used for both local and federated learning. To assess the detection performance, we evaluated the following key metrics, defined as follows:

\begin{equation}
\text{Accuracy} = \frac{TP + TN}{TP + TN + FP + FN}
\end{equation}

\begin{equation}
\text{Precision} = \frac{TP}{TP + FP} \;\;  ; \;\; \text{Recall} = \frac{TP}{TP + FN}
\end{equation}


\begin{equation}
\text{F1-score} = 2 \cdot \frac{\text{Precision} \times \text{Recall}}{\text{Precision} + \text{Recall}}
\end{equation}

In these equations, \(TP\) represents true positives, \(TN\) true negatives, \(FP\) false positives, and \(FN\) false negatives.

\begin{figure}[t]
\centering
\includegraphics[width=1.1 \linewidth]{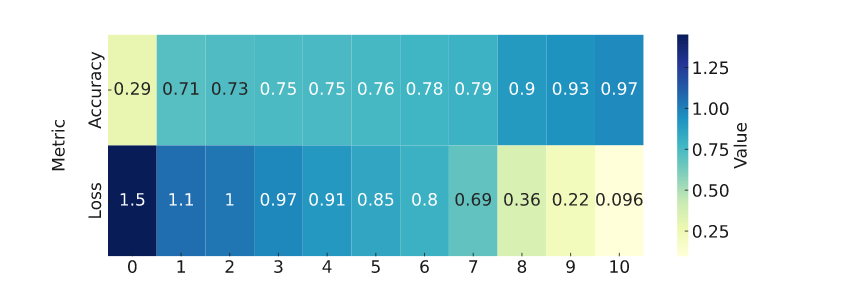}
\caption{Heatmap of Accuracy and Loss Over Rounds for the multiclassifier model with 3 training clients}
\label{fig:heatmap}
\end{figure}

\begin{figure}[t]
\centering
\includegraphics[width=1.1 \linewidth]{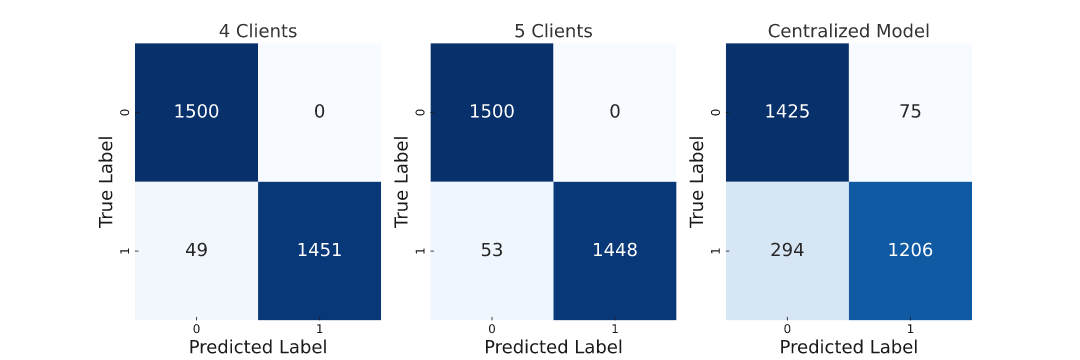}
\caption{Confusion Matrices: DFL (4 and 5 clients) vs. centralized learning}
\label{fig:confusion}
\end{figure}

Figure \ref{fig:heatmap} illustrates the rapid convergence of the model in less than 10 training rounds, highlighting the efficiency of our DFL approach. As detailed in Table \ref{table:metrics}, we evaluated the performance of our bot detection model under both centralized and federated learning configurations, using identical setups with 4 and 5 clients. The results demonstrate that the federated learning model consistently outperforms the centralized model. The centralized approach, constrained by data diversity limitations, fails to generalize effectively, thereby restricting its learning capacity. In contrast, our DFL framework, not only achieves superior detection performance but also ensures transparency and upholds data privacy throughout the training process. Finally, the results are further explained by the confusion matrices in Figure \ref{fig:confusion}, which show that the federated learning model achieves higher detection accuracy and significantly fewer misclassifications compared to the centralized model.

\begin{table}[h!]
\centering
\caption{Binary Classification Performance Metrics: DFL (4 and 5 clients)  vs. centralized learning}
\label{table:metrics}
\begin{tabular}{@{}lcccc@{}}
\toprule
\textbf{Configuration}   & \textbf{Accuracy} & \textbf{Precision} & \textbf{Recall} & \textbf{F1 Score} \\ 
\midrule
Centralized Model    & 0.8770            & 0.9500             & 0.8290          & 0.8854            \\ 
\textbf{DFL} - 4 Clients                 & 0.9837            & 1.0000             & 0.9684          & 0.9839            \\ 
\textbf{DFL} - 5 Clients                & 0.9823            & 1.0000             & 0.9659          & 0.9826            \\ 
\bottomrule
\end{tabular}
\end{table}

\begin{table}[th]
\centering
\caption{Binary Classification Results}
\label{tab:accuracybinary}
\begin{tabular}{@{}lcccc@{}}
\toprule
\textbf{Algorithm}                  & \textbf{Accuracy} & \textbf{Precision} & \textbf{Recall} & \textbf{F1-Score} \\ 
\midrule
Random Forest \cite{b1}      & 0.83             & 0.87              & 0.77           & 0.80            \\ 
Gradient Boosting \cite{b1}  & 0.82             & 0.85              & 0.78           & 0.80            \\ 
AdaBoost \cite{b1}           & 0.83             & 0.84              & 0.80           & 0.81            \\ 
\textbf{Our Work}                                   & \textbf{0.99}    & \textbf{1.00}     & \textbf{0.99}  & \textbf{0.99}   \\ 
\bottomrule
\end{tabular}
\end{table}

\begin{table}[th]
\centering
\caption{Multiclass Classification Results}
\label{tab:accuracymulti}
\begin{tabular}{@{}lcccc@{}}
\toprule
\textbf{Algorithm}                  & \textbf{Accuracy} & \textbf{Precision} & \textbf{Recall} & \textbf{F1-Score} \\ 
\midrule
Random Forest \cite{b1}      & 0.77             & 0.77              & 0.77           & 0.75            \\ 
Gradient Boosting \cite{b1}  & 0.76             & 0.76              & 0.76           & 0.74            \\ 
AdaBoost \cite{b1}           & 0.50             & 0.54              & 0.50           & 0.49            \\ 
\textbf{Our Work}                                   & \textbf{0.96}    & \textbf{0.97}     & \textbf{0.97}  & \textbf{0.97}   \\ 
\bottomrule
\end{tabular}
\end{table}

The binary classification results summarized in Table \ref{tab:accuracybinary} show that our DFL approach with 3 clients outperforms the centralized approaches in terms of accuracy, precision, and recall, achieving an F1 score of 0.99 against at most 0.81 for the centralized approaches. Similarly, the multiclass classification results in table \ref{tab:accuracymulti}, demonstrate the adaptability and efficiency of our DFL approach, which outperforms centralized learning in every metric, with an F1 score of 0.97.

\subsection{Blockchain Performance Evaluation}
The proposed BotDetect is tested on a cluster of two servers 'HPE ProLiant XL225n Gen10 Plus'. Each server has two AMD EPYC 7713 64-Core 2GHz processors and 2x256 GB RAM. We implemented the smart contracts using Solidity and used Geth\footnote{\url{https://geth.ethereum.org}} to run our blockchain. To evaluate the on-chain performance of BotDetect, we run tests using Hyperledger Caliper\footnote{\url{https://github.com/hyperledger/caliper-benchmarks}}. We conduct benchmarks on the two main functions \texttt{submitUpdate} and \texttt{aggregateModel} for a fine-grained assessment of our framework’s performance. We first run a local blockchain network of 16 validators using QBFT as consensus and deploy our main contract. Next, we tailor our benchmarking workloads to capture the system performance associated with each function. The process involves submitting multiple transactions using different send rates: $40-550$ transaction per second (TPS) to measure the throughput and latency of the framework.

\begin{figure}[t]
    \centering
    \hspace{-0.2in}
    \subfloat[\textit{Block Time = 1s}]{
        \includegraphics[scale=0.32]{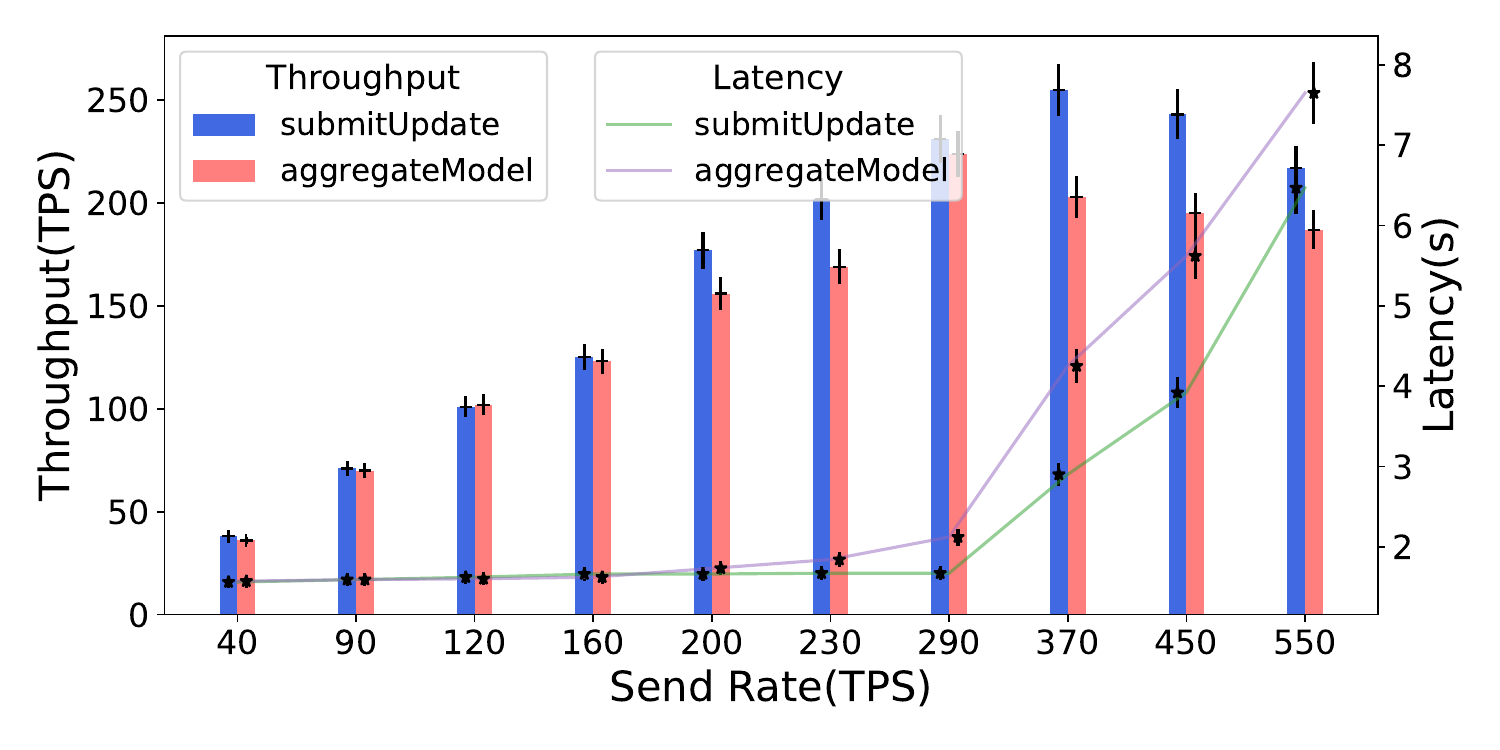}
        \label{fig3a}
    }
    \hfill
    \subfloat[Block Time = 3s ]{
        \hspace{-0.2in}
        \includegraphics[scale=0.32]{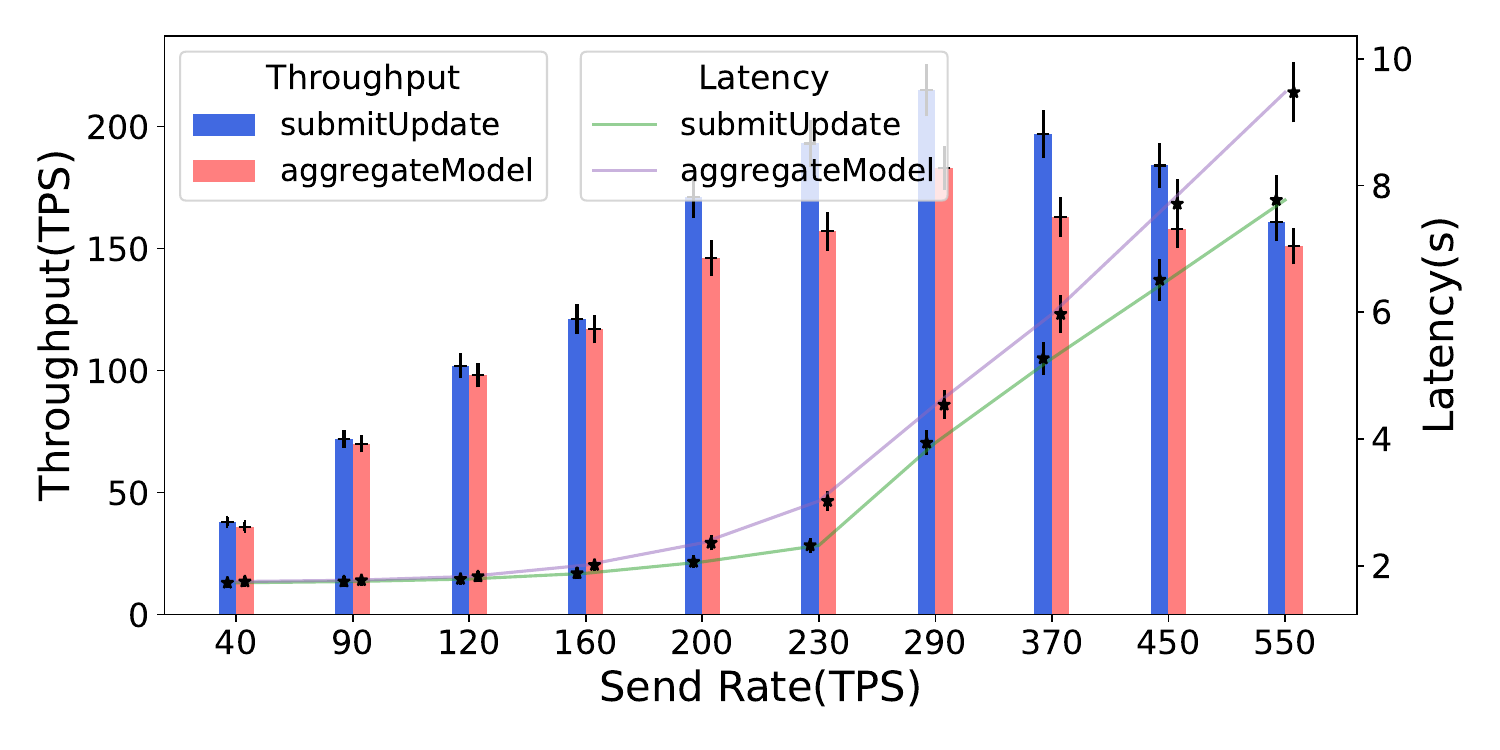}
        \label{fig3b}
    }
    \caption{Latency and Throughput comparison under two
workload types and two block times}
    \label{fig3}
\end{figure}

Figures \ref{fig3a} and \ref{fig3b} show the latency and throughput of BotDetect under two different network configurations (block times=1, 3s). Initially, the pattern is quite clear for both block times: throughput and latency increase as the transaction send rate increases. The \texttt{submitUpdate}, which is lighter than \texttt{aggregateModel} function ($K=5$), achieves the highest maximum throughput of 255$\pm$10 TPS with a send rate of 370 TPS and a block time of 1s. This performance exceeds that of \texttt{aggregateModel} mainly due to the additional compute and memory requirements. 
Additionally, even with a longer block period of 3s, which may be necessary for a wider geographical distribution, the achieved average throughput of 190$\pm$10 TPS is sufficient for BotDetect to operate efficiently.

\section{Conclusion} \label{CON}
In this paper, we presented BotDetect, a decentralized federated learning (DFL) approach for financial bot detection in EVM blockchains. The proposed framework leverages federated learning, orchestrated by smart contracts, to detect malicious bot behavior while preserving privacy and adapting to the decentralized nature of blockchain networks. BotDetect addresses the limitations of centralized approaches by allowing each participating node to train neural network models locally, using data collected from transaction' history. We use on-chain aggregation of model updates through a permissioned consensus mechanism to maintain integrity and decentralization. This design also allows the model to capture complex and evolving bot behavior without requiring direct data exchange between nodes. Experimental results show that our DFL framework achieves high detection accuracy while maintaining efficiency and robustness, providing an effective solution for bot detection in financial blockchains.

For future work, L2 scaling solutions such as zkRollups \cite{b19} can be considered to increase on-chain performance while maintaining greater security through ZKPs.




\end{document}